\documentclass[a4paper,11pt]{article}
\pdfoutput=1 

\usepackage{jinstpub} 
\usepackage{multirow}

\title{LArPix: Demonstration of low-power 3D pixelated charge readout
  for liquid argon time projection chambers}

\author[a,1]{D.~A.~Dwyer,\note{Corresponding author.}}
\author[a]{M.~Garcia-Sciveres,}
\author[a]{D.~Gnani,}
\author[a]{C.~Grace,}
\author[b]{S.~Kohn,}
\author[b]{M.~Kramer,}
\author[a]{A.~Krieger,}
\author[a]{C.~J.~Lin,}
\author[a,b]{K.~B.~Luk,}
\author[b]{P.~Madigan,}
\author[a]{C.~Marshall,}
\author[a,b]{H.~Steiner,}
\author[a]{T.~Stezelberger}

\affiliation[a]{Lawrence Berkeley National Laboratory,\\Berkeley, CA, USA}
\affiliation[b]{University of California -- Berkeley,\\Berkeley, CA, USA}

\emailAdd{dadwyer@lbl.gov}

\abstract{
  We report the demonstration of a low-power pixelated readout system
  designed for three-dimensional ionization charge detection and
  digital readout of liquid argon time projection chambers
  (LArTPCs)\@.
  Unambiguous 3D charge readout was achieved using a custom-designed
  system-on-a-chip ASIC (LArPix) to uniquely instrument each pad in a
  pixelated array of charge-collection pads.
  The LArPix ASIC, manufactured in 180~nm bulk CMOS, provides
  32 channels of charge-sensitive amplification with self-triggered
  digitization and multiplexed readout at temperatures from 80~K to
  300~K\@.
  Using an 832-channel LArPix-based readout system with 3~mm spacing
  between pads, we demonstrated low-noise ($<$500 e$^-$ RMS equivalent
  noise charge) and very low-power ($<$100\,$\mu$W/channel) ionization
  signal detection and readout\@.
  The readout was used to successfully measure the three-dimensional
  ionization distributions of cosmic rays passing through a LArTPC,
  free from the ambiguities of existing projective techniques.
  The system design relies on standard printed circuit board
  manufacturing techniques, enabling scalable and low-cost production
  of large-area readout systems using common commercial facilities.
  This demonstration overcomes a critical technical obstacle for
  operation of LArTPCs in high-occupancy environments, such as the
  near detector site of the Deep Underground Neutrino Experiment
  (DUNE)\@.
}


\arxivnumber{1808.02969} 

\begin{document}
\maketitle
\flushbottom

\section{Introduction}
\label{sec:intro}

Since its invention in 1978~\cite{Nygren:1978rx}, the time-projection
chamber (TPC) has been a prominent particle tracking detector for many
physics experiments~\cite{Olive:2016xmw}.
Energetic particles ionize the bulk detector medium, usually gas or
liquid, liberating electron-ion pairs.
An externally-applied electric field drifts the electrons to a
charge-collection region of the detector, also referred to as the
anode.
The 2D position of the charge arriving at the anode, combined with a
measurement of the drift time, determines the full 3D spatial
distribution of ionization within the TPC\@.
This results in a fully-active tracking detector with homogeneous
characteristics, with no need for instrumentation in the bulk volume.
The ability to drift electrons in cryogenic noble liquids enables a
TPC to function as both a tracking detector and a calorimeter.
The ability to construct large ($>$100~ton) liquid argon
time-projection chambers (LArTPCs) at a reasonable cost makes them
attractive for detecting weakly-interacting
neutrinos~\cite{Rubbia:1977zz, Chen:1976pp, Willis:1974gi}, and was
realized by the ICARUS experiment~\cite{Amerio:2004ze}.
Consequently, large LArTPCs are particularly interesting to the
current international program in neutrino
oscillation~\cite{Anderson:2012vc, Acciarri:2016smi,
  Antonello:2015lea, Acciarri:2016crz}.
%

%
The typical charge readout technique for large LArTPCs uses a stack of
planes of closely-spaced parallel wires at the anode.
The drifting electrons induce signals on nearby wires, and hence each
plane provides an image of the charges in two dimensions: the spatial
position of each wire versus drift time.
Multiple wire planes can provide stereoscopic views that help estimate
the location of ionization in three dimensions, but not without
ambiguity when many wires receive simultaneous signals.
Particle tracks or showers which are oriented nearly coplanar with the
wire readout are particularly difficult to reconstruct.
These ambiguities are exacerbated in high-intensity environments where
the signals from multiple unrelated particle interactions overlap
within the TPC volume, such as the proposed near-site detector of the
Deep Underground Neutrino Experiment (DUNE)~\cite{Acciarri:2016crz}.
Recent experience with wire-based readout of large LArTPCs demonstrate
the challenges not only of signal analysis, but also of noise and
mechanical
fragility~\cite{Acciarri:2017sde,Adams:2018dra,Adams:2018gbi,Qian:2018qbv}.
%

%
Unambiguous 3D imaging of LArTPC charge signals is possible using a
readout system based on a pixelated array of charged-sensitive pads.
This technology is common in gaseous TPCs~\cite{Alme:2010ke,
  Anderson:2003ur, Olive:2016xmw}, but has not until recently been
considered viable for LArTPCs\@.
Much of the difficulty stems from the substantial increase in the
number of electronics channels required for pad readout compared to
wire readout.
Pad-based systems usually locate the front-end electronics close to
the sensor plane to allow digital multiplexing to reduce the signals
to a manageable number of readout channels.
For such a system to function in a LArTPC, requirements beyond those
for wire-based electronics must be met.
In particular, these electronics must be capable of operating in
liquid argon (87~K) and must consume very little power
($<$100~$\mu$W/channel) to avoid excessive heating of the liquid
argon.
Absent electronics that meet these requirements, recent demonstrations
of pad readout in LArTPCs adapted existing wire-based electronics by
combining the signals from tens to hundreds of pads into a single
front-end electronics channel~\cite{Asaadi:2018oxk}.
To mitigate the ambiguities from this arrangement, additional
front-end channels were used to instrument complementary ``regions of
interest'' of the anode.
Comparison of the multi-pad signals with the region-of-interest
signals could partially resolve ambiguities.
These results highlight the importance of electronics that can provide
independent readout of each pad, enabling a true 3D LArTPC\@.
Here we report the demonstration of a pixelated readout system that
provides true 3D readout for LArTPCs\@.
The specifications and design of the readout are discussed in
section~\ref{sec:description}.
Evaluation of the readout at both room and cryogenic temperatures is
given in section~\ref{sec:evaluation}.
The demonstration of true 3D imaging of cosmic ray muon tracks in
LArTPCs is presented in section~\ref{sec:demonstration}.
Concluding remarks are given in section~\ref{sec:discussion}.

\section{Readout Description}\label{sec:description}

The LArPix readout system is designed to detect and acquire signals
from minimum-ionizing particles (MIPs) in LArTPCs\@.
Liquid argon has a density of 1.395~g/cm$^3$ and a boiling point of
87.3~K at standard atmospheric pressure.
The most probable energy loss for minimum-ionizing muons in argon at
this density is 1.66~MeV/cm\@.
The mean energy required to produce an electron-ion pair in liquid
argon is 23.6~eV~\cite{Miyajima:1974zz, PhysRevA.10.1452}.
At a TPC drift field of 500~V/cm, $\sim$30\% of the electrons are lost
to prompt recombination with ions at MIP ionization
densities~\cite{Amoruso:2004dy, Acciarri:2013met}.
This results in $\sim$5000 free electrons per mm of muon track.
While additional losses to impurities (e.g. O$_2$, H$_2$O) can further
reduce signal amplitude, recent detectors have demonstrated electron
lifetimes in excess of 3~ms which alleviates this
concern~\cite{Antonello:2014eha, Acciarri:2016smi}.
Pad spacings in the range of 3~mm to 5~mm are of interest for large
LArTPCs\@.
At coarser spacings particle identification is
hampered~\cite{Acciarri:2016sli}, while electron diffusion during
drift reduces the advantages of finer spacings.
The typical signal amplitude per pad depends on the density of pads,
with pads at smaller spacings (i.e. higher densities) collecting fewer
electrons per pad.
The typical signal from minimum-ionizing muons at 3~mm spacing is
roughly 15,000 electrons per pad, although the specific signal
amplitudes depend on a variety of factors (e.g. track inclination
relative to the readout, geometric overlap of the track with a given
pad, electron diffusion during drift).
A total system noise specification of 500 electron equivalent noise
charge (ENC) corresponds to a 30:1 signal-to-noise ratio (SNR) for
minimum-ionizing muon signals.
Given an electron drift speed of 1.6~mm/$\mu$s at a drift field of
500~V/cm~\cite{Walkowiak:2000wf}, a time resolution of approximately
2~$\mu$s yields a spatial resolution in the drift direction comparable
to the pad spacing.
Pad densities range from 40k to 111k per square meter at the pad
spacings of interest, as shown in Table~\ref{tab:power_requirements}.
The heat load on large LArTPCs is dominated by the heat flow through
the detector cryostat walls, which is generally on the order of
10~W/m$^{2}$\@.
The total area of the anodes for these LArTPCs is typically 0.5 to 1
times the inner surface area of the cryostat.
If the power density of the readout is kept under 10~W/m$^{2}$, then
the total heat load from the readout is less than the heat flux
through the cryostat walls and therefore manageable using existing
cooling techniques.
This limits the power consumption per electronics channel to less than
90 to 250~$\mu$W depending on pad density, as shown in
Table~\ref{tab:power_requirements}.
Detrimental effects from localized boiling of liquid argon can place
additional requirements on local heat density.
Argon gas bubbles within the drift field region of the LArTPC can
result in electrostatic breakdown, while bubbles near the sensitive
analog inputs to the electronics can produce spurious signals.
Existing LArTPCs have been operated with wire-based readout
electronics that have local heat densities of $\sim$0.1~W/cm$^2$,
setting a general target.
The observation and suppression of bubbling during operation of our
pixelated readout system is discussed in section~\ref{sec:evaluation}.

\begin{table}[!htb]
  \begin{center}
    \caption[Pad Densities and Power]{Pad densities and limits for
      average power consumption versus pad spacing.  Average power
      limits are determined assuming a maximum power density of
      10~W/m$^{2}$.\hfill\label{tab:power_requirements}}
    \begin{tabular}{crr}
      \hline
      \hline
      Pad Spacing &  Pad Density & Average Power, Upper Limit \\
      (mm)    &  (m$^{-2}$)  & ($\mu$W/channel) \\
      \hline
      3 & 111,111 &   90 \\
      \hline
      4 &  62,500 &  160 \\
      \hline
      5 &  40,000 &  250 \\
      \hline
      \hline
    \end{tabular}
  \end{center}
\end{table}

%
We designed and assembled a pixelated readout system to meet these
requirements.
Even in high-intensity environments, the O(0.1~cm$^2$)
area of each pad results in an expected signal rate of $\ll$1~Hz per
pad.
These low rates enable a design where much of the electronics is
dormant, expending little power, until a signal occurs.
The LArPix application-specific integrated circuit (ASIC) follows this
approach.
Each of the 32 channels on a LArPix chip functions as an independent
self-triggering signal processor, responsible for amplification,
digitization, and multiplexed readout.
The latter stages of digitization and readout only occur when the
signal on a pad exceeds a configurable threshold, resulting in a
dramatic reduction of the power consumed by the digitizer and data I/O
system during the long intermissions between signals of interest.

\begin{figure}[!htb]
\centerline{ 
  \includegraphics[width=\columnwidth]{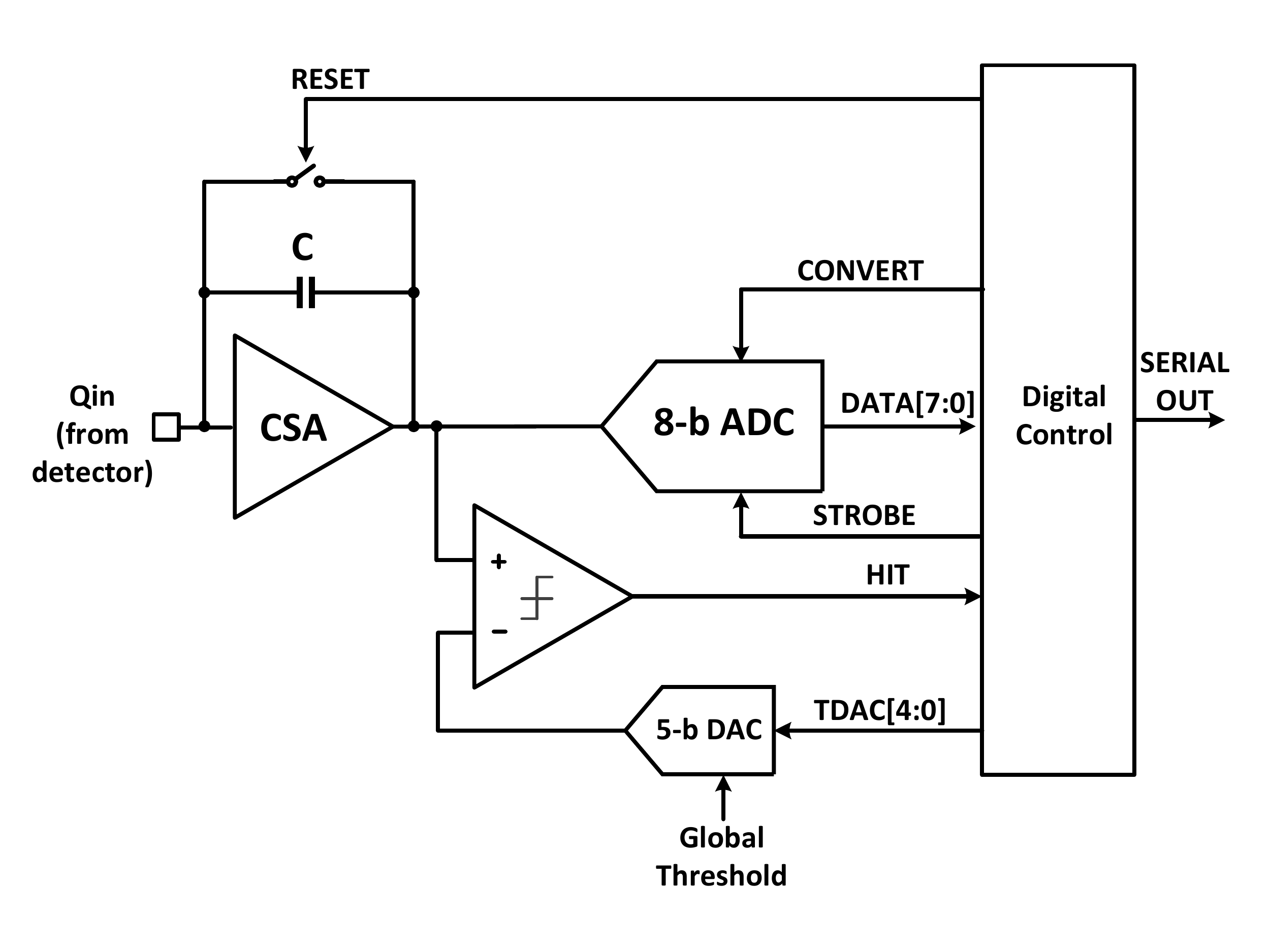}
}
\caption{Block diagram of the LArPix application-specific integrated
  circuit (ASIC).\hfill
\label{fig:larpix_concept}}
\end{figure}

A block diagram of the LArPix design is shown in
Fig~\ref{fig:larpix_concept}.
The signal from each pad is input to a pulsed-reset inverting
charge-sensitive amplifier (CSA) with a rise time of 45~ns and a gain
of 25~mV/fC (4~$\mu$V per electron)\@.
With a quiescent output of 0.55~V at 87~K and a saturation voltage
just under 1.8~V, this provides a dynamic range of 1.2~V or
equivalently 3$\times$10$^5$ electrons at this gain.
The CSA is pulsed-reset with no continuous feedback, providing an
effectively instantaneous peaking time.
As electrons collect on the pad, the voltage at the output of the CSA
grows until it exceeds the threshold of a self-timed discriminator
with a latency of 30~ns\@.
A trigger from the discriminator initiates digitization of the CSA
output using an 8-bit successive approximation register (SAR)
analog-to-digital converter (ADC)\@.
After conversion, the digital control then resets the CSA in
preparation for subsequent signals.
The entire conversion and reset cycle requires 11 clock cycles, or
2.2~$\mu$s with a default 5~MHz system clock.
The ASIC is designed to support a system clock in the range of 2.5 to
20~MHz, providing a minimum time between successive ADC samples of
0.55 to 4.4~$\mu$s\@.
A brief deadtime of three clock cycles occurs at the end of each
triggered cycle, beginning when the CSA output is sampled by the
digitizer and ending when the CSA recovers from reset.
Following each signal, the digital control writes a 54-bit digital
record of the signal to an internal FIFO memory buffer where it awaits
serial transmission out of the ASIC.
Each hit record contains: a 2-bit record type, an 8-bit chip ID, a
7-bit channel ID, a 24-bit timestamp counter based on cycles of the
system clock, an 8-bit ADC value, a 2-bit FIFO status flag, and a
parity bit.
The FIFO buffer on each ASIC holds up to 2048 digital records awaiting
transmission out on a single data wire using a serial UART-like
protocol at a rate of one bit per system clock cycle (e.g.\ 5~Mb/s
with a 5~MHz system clock).
In a similar fashion, each ASIC is configured by passing 54-bit
configuration digital records on a corresponding data input line.
The data output line of one ASIC can be directly connected to the data
input line of a subsequent ASIC; any input data records not intended
for the current ASIC are passed directly to its output.
This data I/O arrangement, referred to as a daisy-chain, enables
communication and controls up to 256 ASICs (8192 channels) via a
single pair of data input and output wires.
This facilitates the operation of large readout systems ($>$10$^4$
channels) using very few data I/O cryostat penetrations, which is
important for minimizing thermal leakage.
A variety of additional features were included in the ASIC design to
allow for flexible operation and testing, with most controlled by
programmable configuration registers within the ASIC\@.
A few of the more relevant configuration options are:
\begin{enumerate}
%
%
\item A signal can be injected into the CSA input on any channel via
  an integrated pulse generator driven by an 8-bit DAC\@.
\item The CSA analog output from any channel can be connected to a
  line out for direct monitoring.
\item An 8-bit coarse threshold DAC with a range from 0 to 1.8~V sets
  a common discriminator threshold for all 32 channels per ASIC, while
  a fine 5-bit trim DAC accommodates relative adjustments in threshold
  for each channel.
\item The least-significant bit (LSB) of the trim DAC is tunable using
  an external bias resistor; the value used in this demonstration was
  1~mV\@.
\item The ADC LSB can be tuned using external reference voltages, and
  is set to a value of 1~mV for the measurements presented here\@.
\item The ADC offset is also set using an external reference voltage,
  accommodating the expected temperature-dependent shift of the CSA
  output voltage.
%
%
\item The front-end can be programmed to periodically reset without
  digitization, draining off any sub-threshold charge that may have
  accumulated on the channel.
\item A programmable channel mask inhibits self-triggering of selected
  channels.
\item A digital external trigger signal forces digitization and
  readout of channels specified via a separate programmable channel
  mask.
\end{enumerate}

The specifications for the LArPix ASIC are summarized in
Table~\ref{tab:requirements}.
%
%
The ASICs were manufactured in 180~nm bulk CMOS technology at a
commercial foundry.
The ASIC die was 5.25~mm by 6.25~mm\@.
Only two of roughly 50 ASICs has failed to function when tested,
suggesting a high yield.

\begin{table}[!htb]
  \begin{center}
  \caption[ASIC Specifications]{Summary of specifications for the
    LArPix ASIC}\label{tab:requirements}
  \begin{tabular}{lrl}
    \hline
    \hline
    Specification &  Value & Comment \\
    \hline
    Analog inputs & 32 & Single-ended input \\
    Gain & 4~$\mu$V/e$^-$ & Optional high gain mode of 45~$\mu$V/e$^-$ \\
    Noise & 500~e$^-$~ENC &  \\
    Power & $<$100~$\mu$W/channel & \\
    Dynamic Range & 1.2~V & Corresponds to $\sim$3$\times$10$^5$ electrons \\
    ADC Resolution & 8 bits & {\em (See discussion in section~\ref{subsec:issues}.)} \\
    ADC LSB & (tunable) & Default: 2~mV \\
    Threshold range & 0 to 1.8~V & Tunable via global and channel trim DACs\\
    Threshold resolution & $<$1~mV & Tunable via external bias resistor \\
    Channel linearity & 1\% & \\
    Timestamp precision & 1 clock cycle & 200~ns, with a 5~MHz system clock \\
    Minimum resample time & 11 clock cycles & From 0.55~$\mu$s to 4.4~$\mu$s (2.5 to 20~MHz)\\
    Operating temperature & 80 to 300~K & \\
    FIFO event memory depth & 2048 &  \\
    Digital data rate & 5~Mb/s & With a 5~MHz system clock \\
    \hline
    \hline
  \end{tabular}
  \end{center}
\end{table}

\begin{figure}[!htb]
\centerline{ 
  \includegraphics[width=1.0\columnwidth]{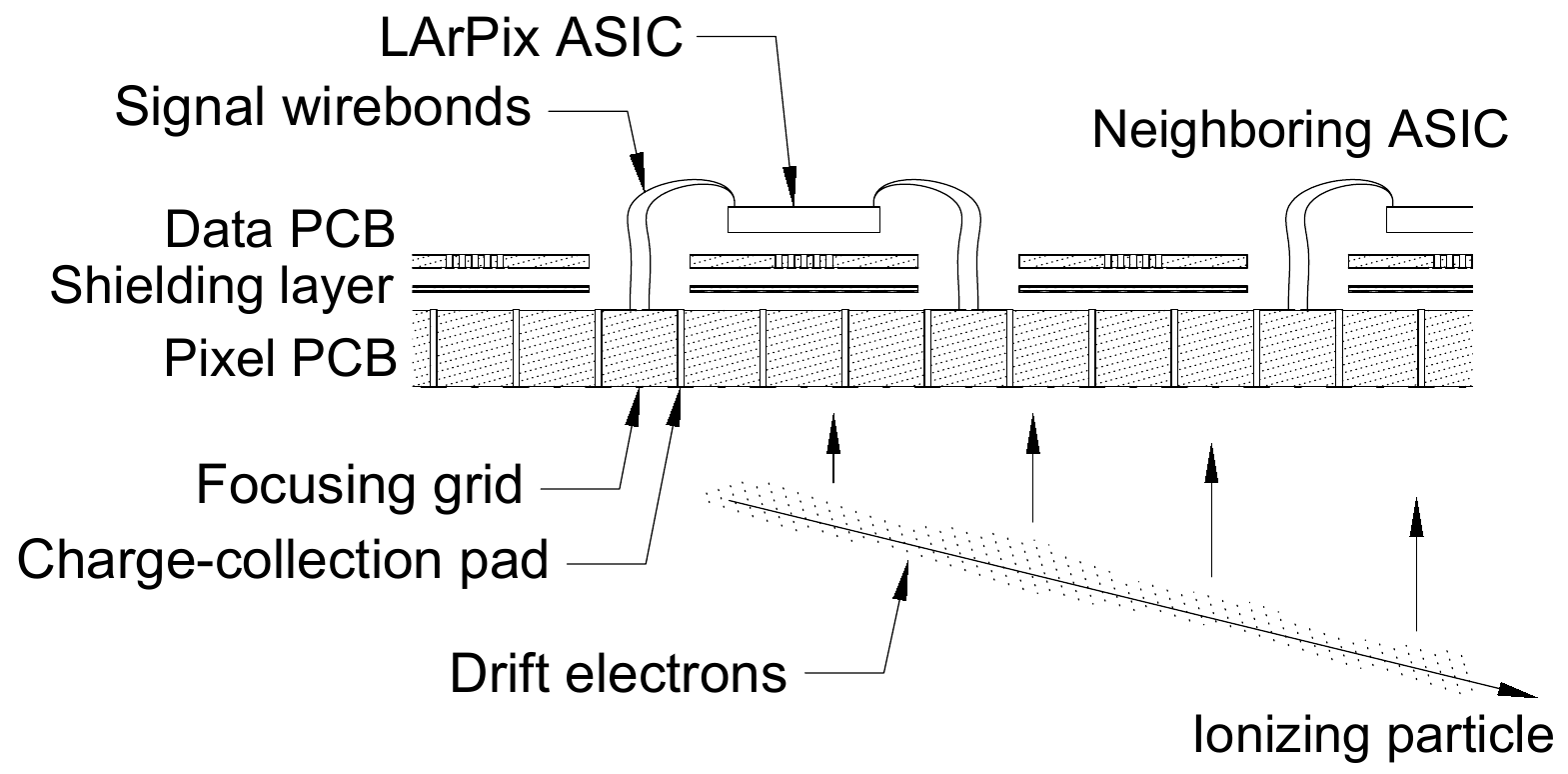}
}
\caption{A cross-section diagram of the readout plane.  Ionization
  electrons were collected on gold-plated copper pads on the Pixel
  PCB\@.  The signals from each pad were transmitted via wire bonds to
  a unique input channel on a 32-channel LArPix ASIC\@.  The ASIC
  amplified, digitized, and multiplexed the digital signals out of the
  system.  The Data PCB provided power as well as data input and
  output routing for the ASIC\@.  The shielding layer reduced the
  cross-talk from the Data PCB digital activity to the Pixel PCB\@.
  \hfill
\label{fig:readout_crosssection}}
\end{figure}

\begin{figure}[!htb]
\centerline{ 
  \includegraphics[width=0.35\columnwidth]{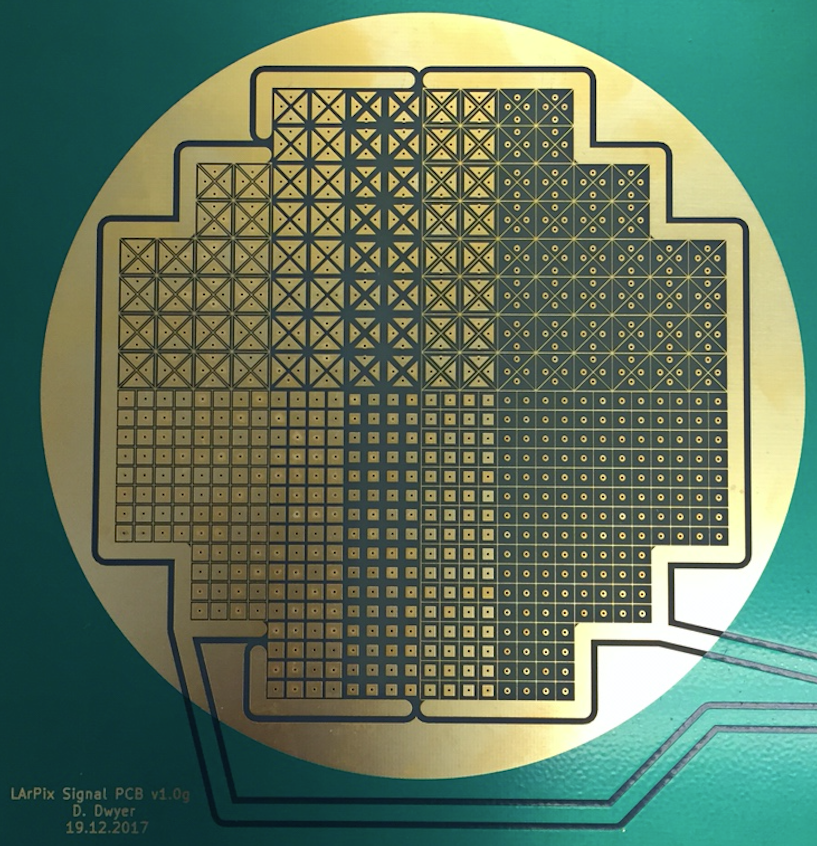}
  \includegraphics[width=0.60\columnwidth]{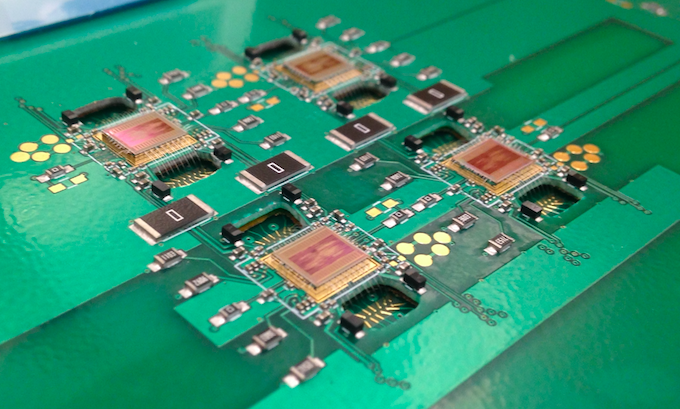}
}
\caption{ {\em Left:} Photograph of the TPC-facing side of a pixelated
  readout system.  A total of 832 pads are etched on a standard
  two-layer circuit board.  Ten different pad configurations are
  included in order to assess their relative performance.  {\em
    Right:} Photograph of the back side of the readout assembly.  Four
  LArPix ASICs are mounted on a two-layer data PCB responsible for
  routing system power and digital communication.  For this readout
  system only 128 of the pads are instrumented, each wire bonded to a
  unique analog input of the ASICs through oblong cavities cut in the
  data PCB\@. \hfill
\label{fig:sensor_assembly}}
\end{figure}

A LArPix-based pixelated readout system was designed to instrument the
10-cm-diameter circular anode of a test LArTPC\@.
A cross-section diagram of the readout assembly design is shown in
figure~\ref{fig:readout_crosssection}\@.
A total of 832 gold-plated copper pads spaced at 3~mm were etched onto
one surface of a standard two-layer 1/8-inch-thick FR-4 circuit board
(see left panel of figure~\ref{fig:sensor_assembly}), here referred to
as the pixel PCB\@.
Two pad shapes (square and triangle) and three pad sizes (large,
medium, and small) were included in this design to facilitate study of
their relative performance.
A focusing grid surrounds the 416 pads on the right half of the pixel
PCB.
This grid is composed of a network of 0.15~mm-wide PCB traces etched
at the boundaries between these pads\@.
Biasing this grid at voltages in the range of $-$100~V to $-$300~V
focuses drifting electrons to increase the charge collection
efficiency of these smaller pads.
A grid of similar design surrounds the left-most 160 large pads, and
is included to study the impact of a grounded mesh separating these
pads.
Vias carry the signal from each pad to the back side of this PCB,
where short (2 to 5~mm) traces route the signals from groups of 16
pads to small 1~mm by 5~mm regions for coupling to the readout
electronics.

Wire bonds connect each signal trace to a unique analog input of a
LArPix ASIC\@.
A separate thin two-layer PCB, here called the data PCB, hosts the
LArPix ASICs and routes the power, clock, and data input and output
necessary for their operation.
A data PCB designed to host four LArPix ASICs, as shown in the right
panel of figure~\ref{fig:sensor_assembly}, was used for most of the
initial characterization.
%
%
Cross-talk between the digital activity on the data PCB and the pads
on the pixel PCB was reduced by a shield layer consisting of a
grounded copper sheet covered by polyamide to avoid electrical contact
to either of the PCBs\@.
Small 3~mm by 6~mm cavities through the data PCB and shield allow wire
bond connections from each LArPix analog input directly to the signal
traces on pixel PCB, as visible in figure~\ref{fig:sensor_assembly}\@.
A custom-designed digital control system, located outside the
cryogenic detector, provides power, reference voltages, a system
clock, digital data input and output, and an integrated DAQ system.
A Digilent Cmod A7 FPGA development board generates the system clock,
and also serves as a data I/O bridge between the real-time environment
of the LArPix ASICs and the DAQ computer.
A miniature Raspberry Pi Zero computer running Linux (Raspbian
Stretch) serves as an integrated DAQ computer and control system.
A light-weight DAQ software architecture, written in Python, provides
a convenient user interface for the control and operation of the
readout system.
Communication with the DAQ computer occurs via standard 802.11 WiFi
techniques, facilitating electrical isolation of the entire TPC
readout system from the external environment.

\section{Performance Evaluation}\label{sec:evaluation}

Initial assessment of the LArPix ASICs was performed using a
128-channel (four-chip) LArPix readout system.
Measurements of basic communication, gain, noise, and power were made
both at room temperature and immersed in a liquid nitrogen (LN) bath
(77~K)\@.
Additional measurements were performed with the LArPix ASIC analog
inputs left unconnected from the charge-collection pads, thereby
assessing the intrinsic characteristics of the chip.
No issues with the basic communication with or configuration of the
ASICs were encountered under any of these conditions.
Reading and writing of all of the internal configuration registers of
the ASIC functioned as designed.
The readout system showed no problems following multiple thermal
cycles between room temperature and immersion in either LN or LAr\@.
The details of specific measurements are discussed below.

\subsection{Leakage Current}

Using the analog monitoring feature of the LArPix ASIC, the CSA output
voltage from each channel was observed to increase linearly with time.
At room temperature, this increase was consistent with a small leakage
current of $\sim$80~fA, or roughly 500 electrons per millisecond, into
the analog input of each channel.
The rising CSA output voltage would eventually exceed the
discriminator threshold, resulting in self-triggering and reset at
rates of a few~Hz per channel.
Enabling the periodic reset feature at rates of 3~kHz to 10~kHz was
sufficient to drain away this slowly-accumulating charge with minimal
residual bias to the CSA output voltage level.
When immersed in LN, the leakage decreased to approximately 80~aA
(500~e$^-$/s)\@.
After bonding the pads to the ASIC channel inputs, the leakage
increased by an amount which varied amongst the channels.
A thorough cleaning of the pads using isopropyl alcohol was sufficient
to suppress this leakage current, which was attributed to
contamination on the surface of the pixel PCB\@.

\subsection{Gain}

\begin{figure}[!htb]
\centerline{ 
  \includegraphics[width=0.8\columnwidth]{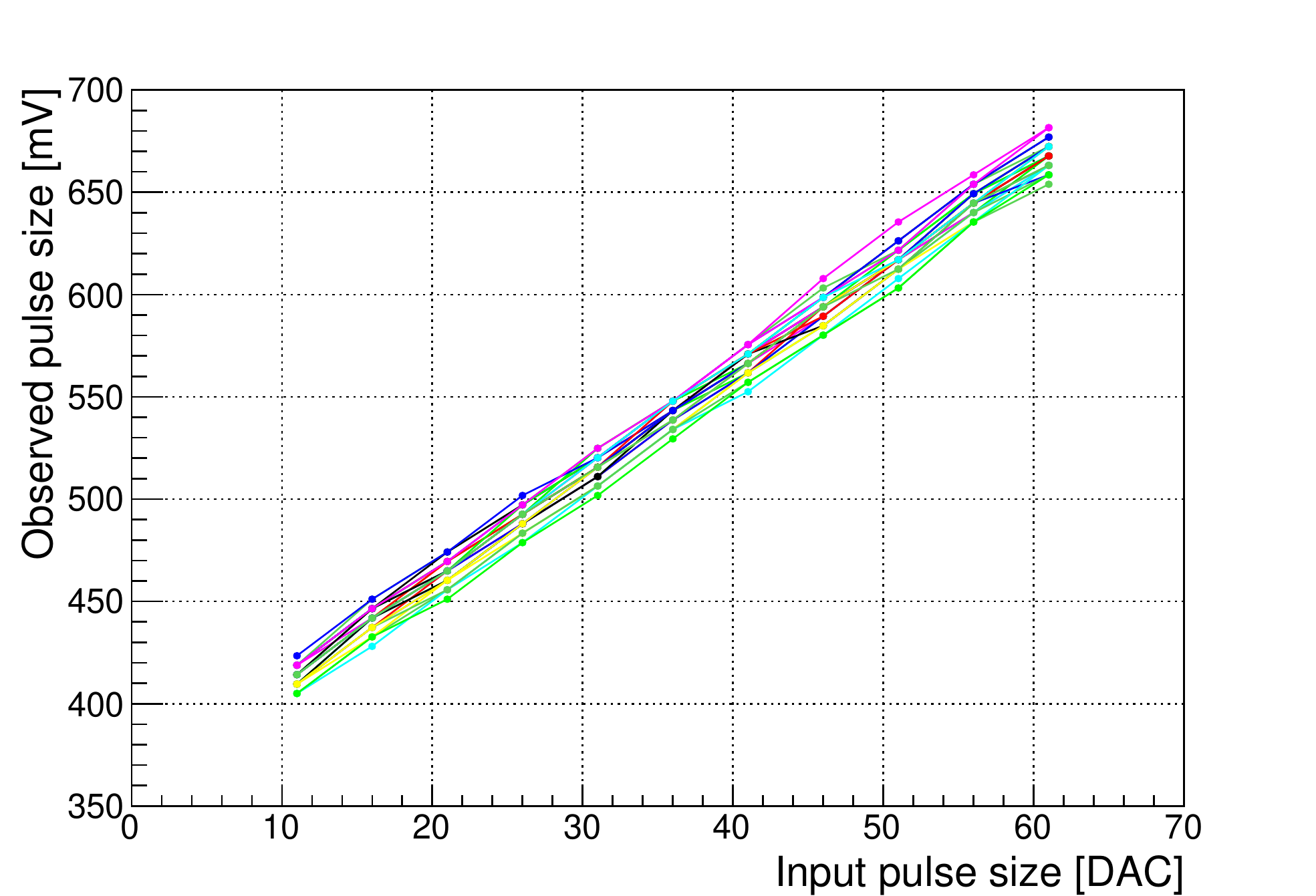}
}
\caption{ Plot of the self-triggered ADC value versus internal pulser
  DAC value for the 32 channels of one LArPix ASIC\@.  The slope of
  each line provides a preliminary assessment of the CSA gain,
  consistent with the design specification of 4~$\mu$V per electron.
  The digital ADC value has been converted to voltage based on an
  initial calibration of the ADC scale.\hfill
\label{fig:gaintest}}
\end{figure}

The built-in pulser was used to evaluate the gain of the CSA\@.
Pulses of increasing amplitude, as programmed using the 8-bit pulser
DAC, were injected into each channel and examined using the analog
monitor feature of the LArPix ASIC.
This same procedure was then repeated, but with the self-trigger
enabled.
The resulting ADC values (calibrated to mV) versus pulser DAC value
are shown in figure~\ref{fig:gaintest}.
All 32 channels of this ASIC showed the expected relationship of
4~$\mu$V per electron, given the approximately 1500 electrons per DAC
count of the internal pulser.
The precision of the internal pulser was not sufficient to assess the
gain linearity at the 1\%-level given by the design specification; a
precise assessment is still pending.
No significant change in the gain was observed when the system was
immersed in LN\@.

\subsection{Pedestal}

\begin{figure}[htb]
\centerline{ 
  \includegraphics[width=0.5\columnwidth]{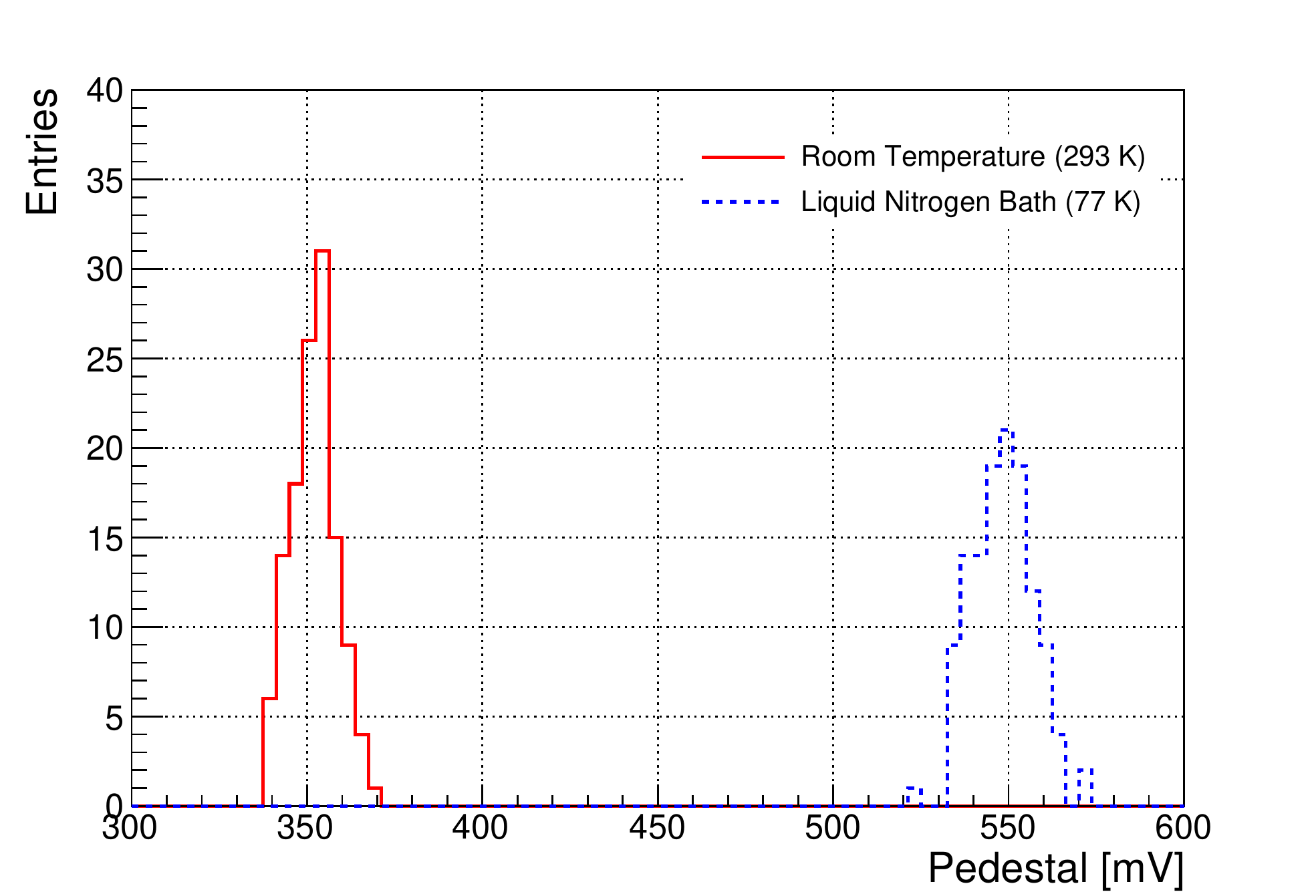}
  \includegraphics[width=0.5\columnwidth]{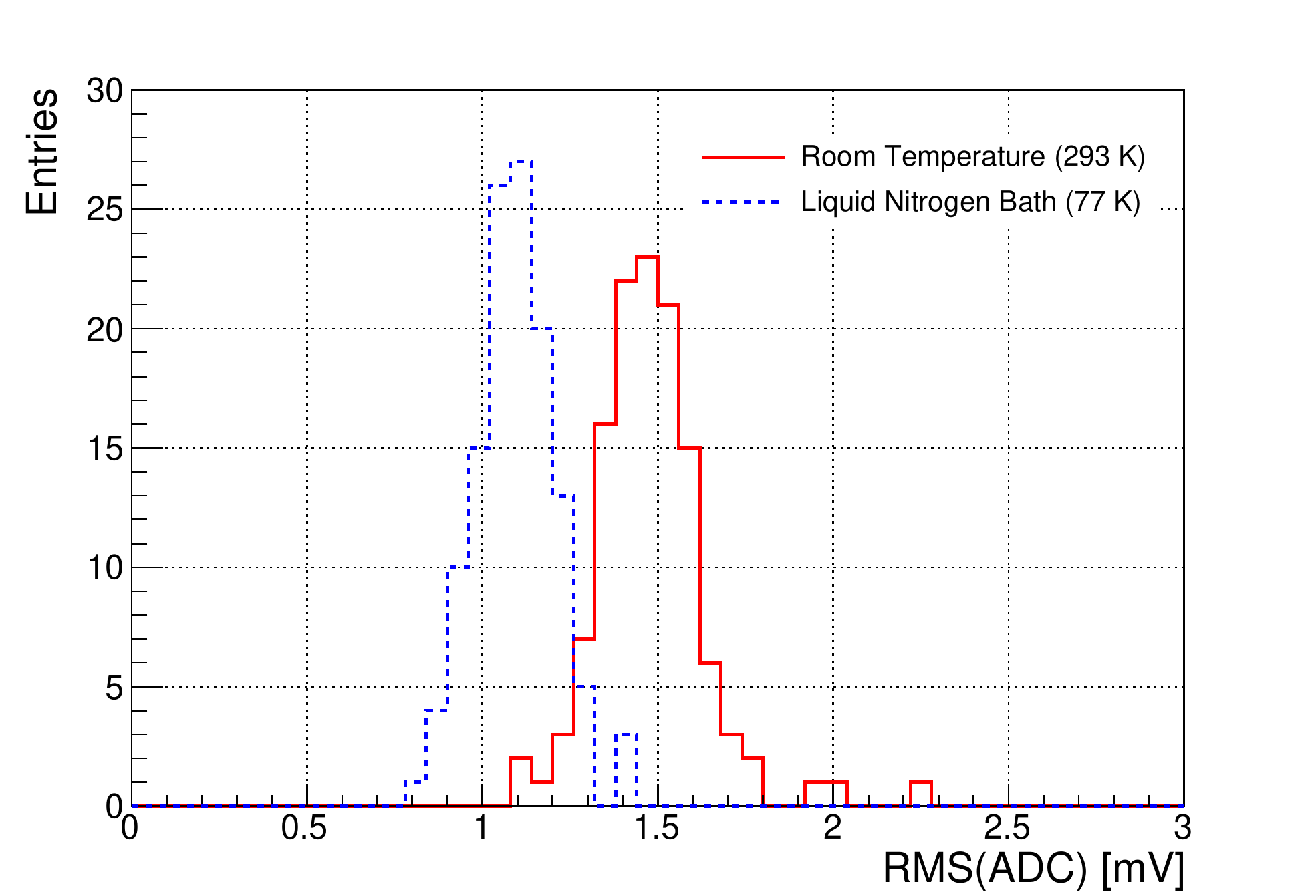}
}
\caption{{\em Left:} Histograms of the pedestal voltages for 128
  LArPix channels at room temperature (solid red line) and when
  immersed in liquid nitrogen (dashed blue line).  The mean voltage is
  350~mV at room temperature.  This increases to 550~mV when the
  system is immersed in liquid nitrogen, which is consistent with the
  expected temperature dependence of the analog front-end.  The spread
  between the channels with the lowest and highest pedestal voltages
  is 30~mV at room temperature, and increases to 50~mV in liquid
  nitrogen.  {\em Right:} Histograms of the noise voltage of 128
  LArPix channels at room temperature (solid red line) and when
  immersed in liquid nitrogen (dashed blue line).  An average noise of
  1.5~mV (375~e$^-$~ENC) is observed at room temperature, and
  decreases to 1.1~mV (275~e$^-$~ENC) in a liquid nitrogen bath.\hfill
\label{fig:noisetest}}
\end{figure}

The mean DC pedestal voltage at the output of the CSA was
approximately 350~mV, as measured using the analog monitoring feature
of the LArPix ASIC\@.
This mean pedestal voltage was consistent with the simulation of the
LArPix front-end.
The pedestal voltages were also confirmed by extrapolation of injected
pulses to zero amplitude (i.e. the y-intercepts of the lines in
figure~\ref{fig:gaintest}), as well as from the distribution of ADC
values obtained via forced digitization when no signal was present on
the channels (see left panel of figure~\ref{fig:noisetest}).
The average pedestal voltage shifted from roughly 350~mV at room
temperature to 550~mV in liquid nitrogen, which was also consistent
with the expected temperature dependence of the front-end.
The spread between the channels with the lowest and highest pedestal
voltages was 30~mV at room temperature, and increased to 50~mV in
liquid nitrogen.
This channel-to-channel variation exceeded the expectations based on
ASIC simulation, and was outside the default range for relative
adjustment of channel thresholds (i.e. 5-bit trim DAC with a 1~mV
LSB)\@.
The trim DAC LSB is tunable through modification of external bias
circuits on the Data PCB, so the LSB will be increased in future
versions.
The impact of insufficient trim range on channel thresholds during
LArTPC operation will be discussed further in
section~\ref{sec:demonstration}\@.

\subsection{Noise}

The noise performance of the system was evaluated using forced
digitization of each channel when no signal was present.
The RMS of the ADC distribution for each channel provided a combined
assessment of the noise from both the CSA as well as the digitizer.
At room temperature, the 128 channels from four LArPix ASICs showed an
average RMS of 1.5~mV, as displayed in the right panel of
figure~\ref{fig:noisetest}\@.
A 1.5~mV RMS implies an equivalent noise charge of roughly
375~e$^-$\@.
When immersed in LN, the noise decreased to roughly 1.1~mV
(275~e$^-$~ENC)\@.
This noise level is consistent with the design specifications shown in
Table~\ref{tab:requirements}, and suggests a potential signal-to-noise
ratio of 55:1 for minimum-ionizing muon tracks.

\subsection{Power}

Three separate supply voltages were used to power the readout system.
The analog supply voltage (VDDA), with an operating range from +1.0~V
to +1.8~V, drove the analog portion of the ASIC (CSA and part of the
digitizer) and also served as a source of bias currents.
The digital core supply voltage (VDDD) powered the digital functions
of the ASIC (digital core and part of the digitizer), and had an
operating range from approximately +1~V to +1.8~V\@.
The digital I/O supply voltage (VDDIO) drove the CMOS data I/O to and
from the ASIC, as well as the ASIC electrostatic discharge (ESD) ring.
By default VDDIO was +3.3~V but had an operating range down to
+1.5~V\@.
Given the low-power design of the readout, we were able to power the
system using three standard AAA batteries.
These batteries were used in combination with three low-dropout (LDO)
linear regulators (Analog Devices LT3080) and three potentiometers
(Bourns 3296) to provide convenient tunable low-noise power supplies
for VDDA, VDDD, and VDDIO\@.
Using three separate supply voltages allowed us to independently
measure and tune the power consumption of each functional component of
the ASIC\@.
We measured the power consumption by placing a digital voltage meter,
in current measurement mode, in series with each of the three supply
voltages: VDDA, VDDD, VDDIO\@.
Table~\ref{tab:powertest} summarizes the results of these measurements
when operating at room temperature and at a clock operating frequency
of 5~MHz\@.
Using an initial configuration of +1.5~V, +1.8~V, and +3.3~V for VDDA,
VDDD, and VDDIO respectively, we measured an average power of
294~$\mu$W per channel.
We found stable operation of the readout system at VDDD values down to
+1.1~V; at lower voltages bit errors began to occur in the data
generated by the system.
The digital I/O functioned correctly down to a VDDIO value of +2.0~V;
below this value our control electronics failed to stably register the
data returned by the readout system\@.
It is possible that the system was still functional at lower VDDIO
supply voltages, but we were unable to record and verify this data.
With these lower digital supply voltages, we found an average power
consumption of 62~$\mu$W per channel for this 128-channel system.
The power use was shared roughly equally across the analog, digital
core, and digital I/O functions.

\begin{table}[!htb]
  \begin{center}
    \caption[Power Consumption Measurement]{Average power consumption
      per channel for a 128-channel readout system at a clock
      operating frequency of 5~MHz\@.  Values are given for two
      configurations: an initial default configuration of the supply
      voltages, and a configuration tuned for low-power
      operation.\hfill\label{tab:powertest}}
    \begin{tabular}{ccrrr}
      \hline
      \hline
      Configuration &  Source & Voltage & Current & Average Power \\
      & & (V)    &  (mA)  & ($\mu$W/channel) \\
      \hline
      \multirow{4}{*}{Default} & VDDA & 1.5 & 2.0 & 24 \\
      & VDDD & 1.8 & 3.6 & 51 \\
      & VDDIO & 3.3 & 8.5 & 219 \\
      & & & & \textbf{Total: 294} \\
      \hline
      \multirow{4}{*}{Low-power} & VDDA & 1.5 & 2.0 & 24 \\
      & VDDD & 1.1 & 2.1 & 18 \\
      & VDDIO & 2.0 & 1.3 & 20 \\
      & & & & \textbf{Total: 62} \\
      \hline
      \hline
    \end{tabular}
  \end{center}
\end{table}

We also examined the variation of the power consumption with readout
activity level.
We lowered the threshold of one channel to the point that the data
transmission through the four chip daisy chain was saturated, reaching
a rate of roughly 10~kHz\@.
This resulted in a $\sim$20\% increase in total system power
consumption.
In a more extreme test, we lowered the thresholds on all channels such
that the digital cores and internal FIFOs of all four chips were
saturated.
For this extreme case, the total power increased by a factor of 4\@.
In the expected regime of operation, with trigger rates of less than
1~Hz per channel, the power increased by only a few percent.
Power consumption did not significantly increase when
charge-collection pads were connected to the ASIC inputs, nor when the
system was immersed in LN.

\subsection{Digital Multiplexing}

\begin{figure}[!htb]
\centerline{ 
  \includegraphics[width=0.7\columnwidth]{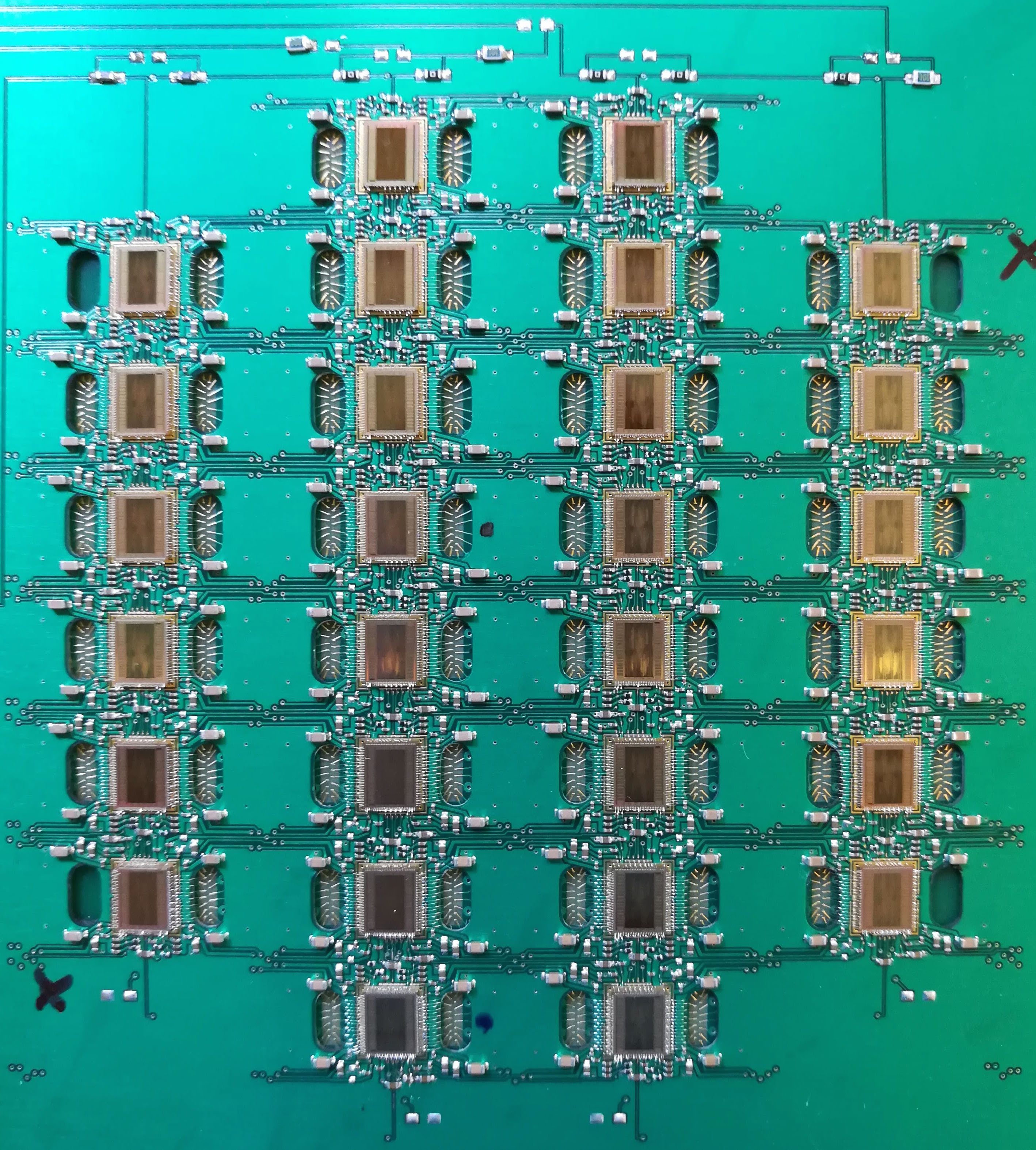}
}
\caption{ Photograph of the digital side of an 896-channel (28-chip)
  readout assembly.  This larger system demonstrated scalability of
  the pixelated readout design, and was used to fully instrument and
  operate a 10-cm-diameter LArTPC\@. \hfill
\label{fig:28chip_sensor}}
\end{figure}

We exercised the digital multiplexing features of the LArPix ASIC
throughout the testing program.
The four chips of the readout system were operated as a single daisy
chain, with all configuration and control of the system requiring only
one data input and one data output line.
By modification of jumpers on the data PCB, we also exercised the data
I/O in configurations of 1, 2, and 3 ASICs in a chain.
An 896-channel daisy chain was exercised following the design and
production of the 28-chip PCB shown in
figure~\ref{fig:28chip_sensor}\@.
This scalable layout was used to instrument all 832 pads of a pixel
PCB (e.g. left panel of figure~\ref{fig:sensor_assembly})\@.
In general, the daisy chain functionality performed as designed and no
obstacle was found that would prevent scaling the digital multiplexing
to the maximum allowed by the LArPix design: 8192 channels per pair of
data I/O lines.

\subsection{Known Issues}\label{subsec:issues}

The following issues were identified during the development of this
readout system.
The default values for the ASIC configuration registers were such that
the self-triggering was active with a low threshold when the system
was initially powered.
This caused saturation of the data output and an initial increase in
power consumption, but was easily resolved by prompt reconfiguration
of the thresholds of all channels\@.
The initial version of the ASIC included a 6-bit SAR ADC design, which
was later enhanced to 8 bits.
During testing, the ADC bits of lowest and highest significance were
found to have fixed values, thus providing only the original 6-bit SAR
performance.
An error in the design was promptly identified via simulation, and the
design was revised to provide full 8-bit performance for the next
version of this ASIC\@.
The timestamps for simultaneous externally-triggered hit records were
found to be offset by 3 clock cycles.
This offset resulted from a relative time delay when writing multiple
simultaneous hit records to the FIFO\@.
Synchronizing the timestamp with the trigger, instead of with the hit
record entry to the FIFO, would resolve this minor issue.
None of these issues presented a serious technical obstacle to
operation and testing of the system, but their resolution must await
production of a revised LArPix ASIC\@.

\section{Demonstration of 3D Imaging of Cosmic Rays}\label{sec:demonstration}

\begin{figure}[!htb]
  \centerline{
    \includegraphics[width=0.67\columnwidth]{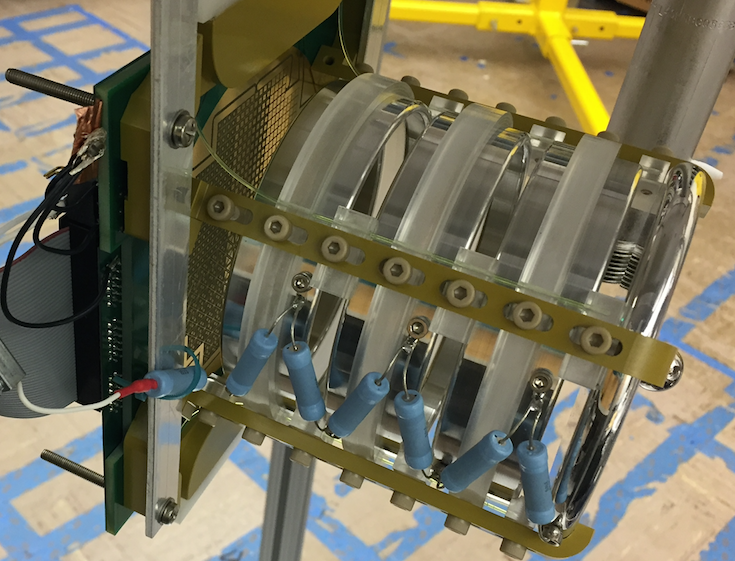}
    \includegraphics[width=0.325\columnwidth]{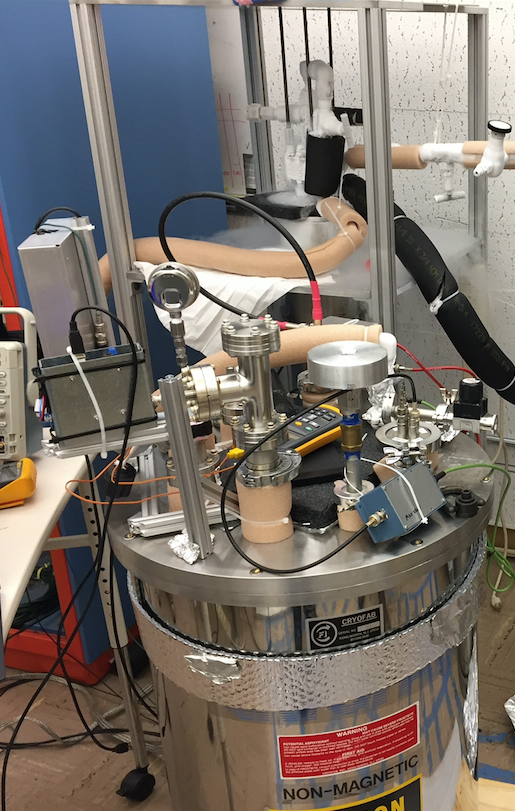}
}
\caption{ {\em Left:} Photograph of the 10-cm-drift TPC with a 3D
  readout system installed. {\em Right:} Photograph of the high-purity
  liquid argon system during operation.  The TPC is installed in the
  cryostat (foreground).  A single-pass of research-grade liquid argon
  through the LAr purifier (background) is sufficient to achieve
  electron lifetimes in excess of 100~$\mu$s\@. \hfill
\label{fig:10cmTPC}}
\end{figure}

%
A small 10-cm-diameter by 10-cm-long LArTPC was used for an initial
demonstration of MIP track detection using the described pixelated
readout system.
This TPC, shown in figure~\ref{fig:10cmTPC}, was designed and built by
collaborators at the University~of~Bern\@.
We instrumented this TPC first with a 4-chip (128-channel) LArPix
system and subsequently with a 28-chip (832-channel) system.
We installed this TPC in an existing 20-liter high-purity LAr system
which had been used for development and testing of liquid argon purity
monitors~\cite{LuciesThesis}\@.
We enhanced this system with
a custom-designed capacitive liquid argon level sensor to monitor the
LAr level during TPC operation,
a muon telescope to provide a tagged sample of cosmic rays passing
through the TPC,
a resistive heater to facilitate the return to room temperature after
operation,
and two resistive temperature sensors for monitoring the cryostat
during filling and purging as well as for monitoring the readout
system temperature during operation.
We arranged the system such that the cryostat lid served as the star
ground point for all components internal to the cryostat, including
the LArPix charge readout system.
We also installed low-pass RC filters at the cathode high voltage and
bias grid voltage feedthroughs on the cryostat lid.
With this configuration, the noise level was found to be roughly
1.5~mV at room temperature and 1.1~mV in liquid argon, consistent with
that observed in stand-alone tests.

\begin{figure}[!htb]
  \centerline{
    \includegraphics[width=\columnwidth]{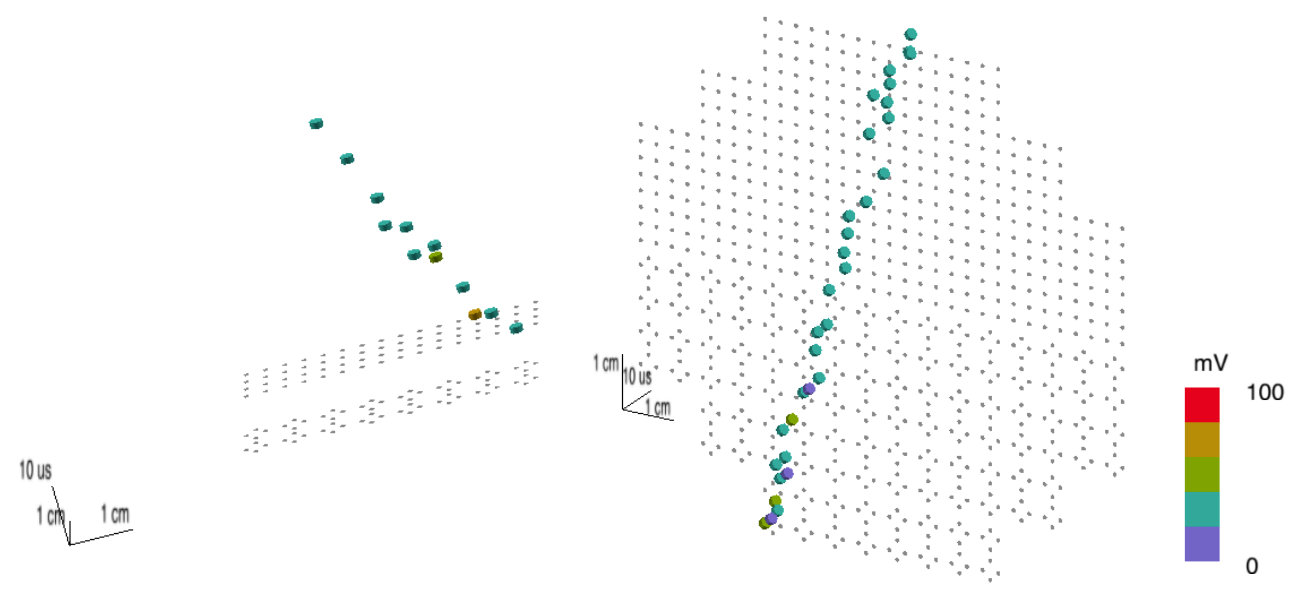}
  }
\caption{ {\em Left:} 3D image of a cosmic ray muon which passed at a
  steep angle relative to the face of a 128-channel readout system.
  {\em Right:} A similar image of a cosmic ray muon passing nearly
  coplanar to a fully-instrumented 832-channel readout system.  Both
  figures display the raw data generated by the readout system; no
  filtering or signal enhancement has been applied.  \hfill
\label{fig:tracks_10cm}}
\end{figure}

The apparatus was filled with high-purity liquid argon using a
single-pass molecular sieve and activated copper purifier to remove
O$_2$ and H$_2$O\@.
The TPC cathode was brought to $-$2~kV using a Canberra 3002D
high-voltage supply, providing a drift field of 200~V/cm to the TPC,
while the readout anode operated at ground.
A separate supply (Ortec 710) was used to bias the focusing grid at
$-$200~V, ensuring efficient electron collection by the smaller-sized
pads.
The thresholds of each channel were adjusted so that the hit rates
from noise were less than 1~Hz per channel, and self-triggered data
was collected.
Upon examination of this self-triggered data, the signals from cosmic
ray muons were easily identified as bursts of hit records arriving
within $\sim$100~$\mu$s of each other.
Figure~\ref{fig:tracks_10cm} shows two 3D images of cosmic ray raw
data, the left panel taken with the 128-channel system and the right
panel with the 832-channel system.
Each large colored point in this figure represents one hit record,
where the color provides a rough estimate of the charge (with blue to
green approximately covering the range of 10k to 20k electrons).
Each small gray point represents the position of a single pad from the
pixel PCB shown in figure~\ref{fig:sensor_assembly}\@.
In this display, the sensor face of the readout system is shown in a
tilted perspective and each hit is displaced from the face using the
relative offset of each hit timestamp.
A scale of 1~mm/$\mu$s was used for this offset, consistent with the
expected electron drift velocity in liquid argon at an applied field
of 200~V/cm\@.

\begin{figure}[!htb]
  \centerline{
    \includegraphics[width=0.6\columnwidth]{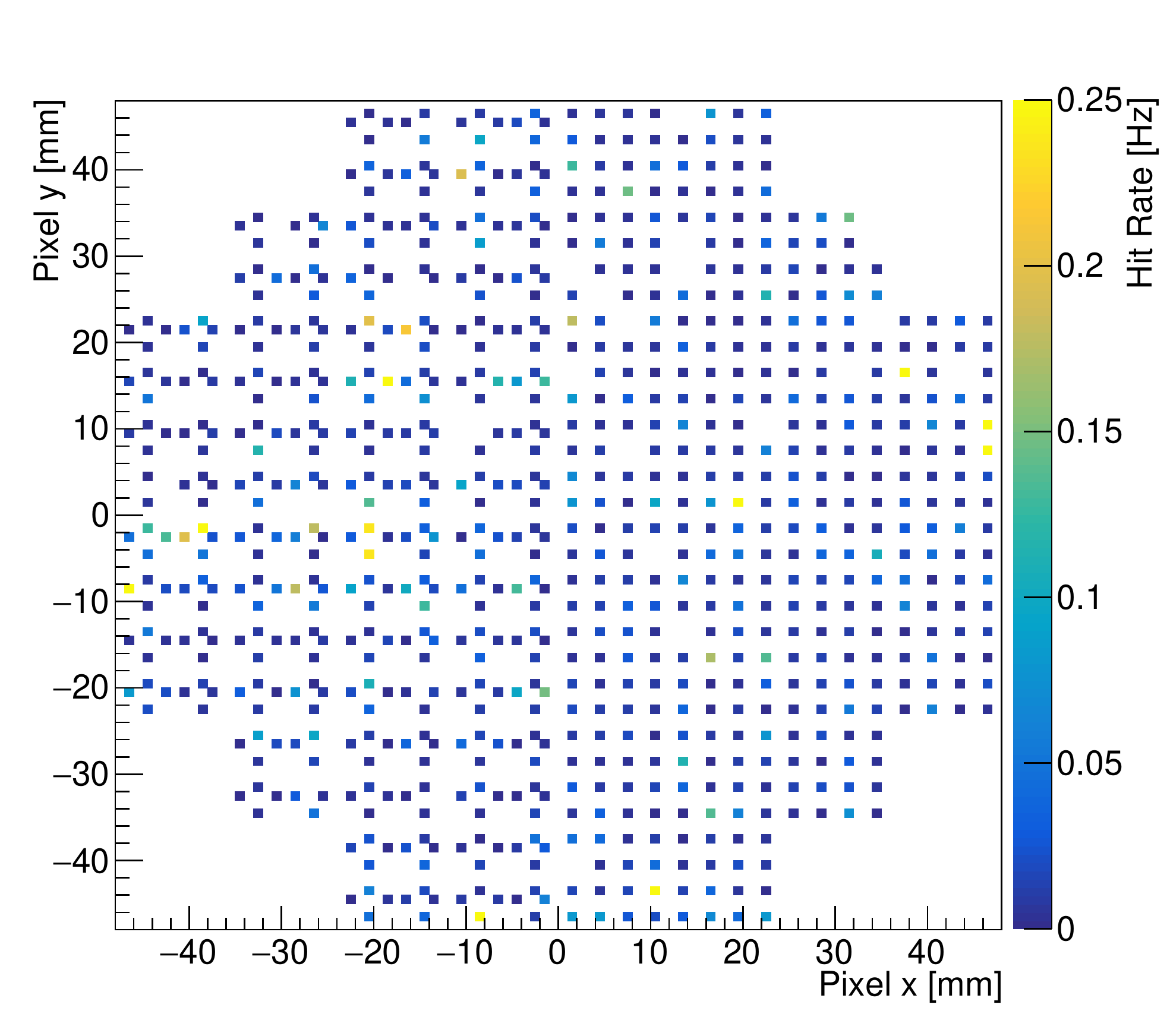}
  }
\caption{ Channel self-trigger rates for an 832-channel pixelated
  readout system with 3~mm spacing.  The majority of channels have low
  rates ($\sim$0.02~Hz) that are dominated by the cosmic ray signals
  of interest.  A few channels have low thresholds and high rates
  ($>$0.05~Hz) that are primarily noise.  A few-percent of channels
  were disabled due to high noise rates, and are visible as gaps in
  the pixelated array. \hfill
\label{fig:28chip_results}}
\end{figure}

Figure~\ref{fig:28chip_results} shows the average self-triggered rate
of each channel taken with the 28-channel system.
The 50~mV spread in the pedestal voltages among the channels exceeded
the range over which we could tune relative differences in threshold
among the channels.
The self-triggering of the system was optimized by disabling a few
percent of channels with the highest pedestal voltages (visible as
gaps in figure~\ref{fig:28chip_results}), and the thresholds of the
remaining channels were tuned to achieve sensitivity to MIP signals.
After tuning, the majority of channels had self-trigger rates of
roughly 0.02~Hz, consistent with the estimated rate from cosmic ray
muons.
A few remaining channels with high pedestal voltages were kept active
and gave manageable 0.05~Hz to 1~Hz trigger rates from noise, as shown
in figure~\ref{fig:28chip_results}\@.

\begin{figure}[!htb]
  \centerline{
    \includegraphics[width=\columnwidth]{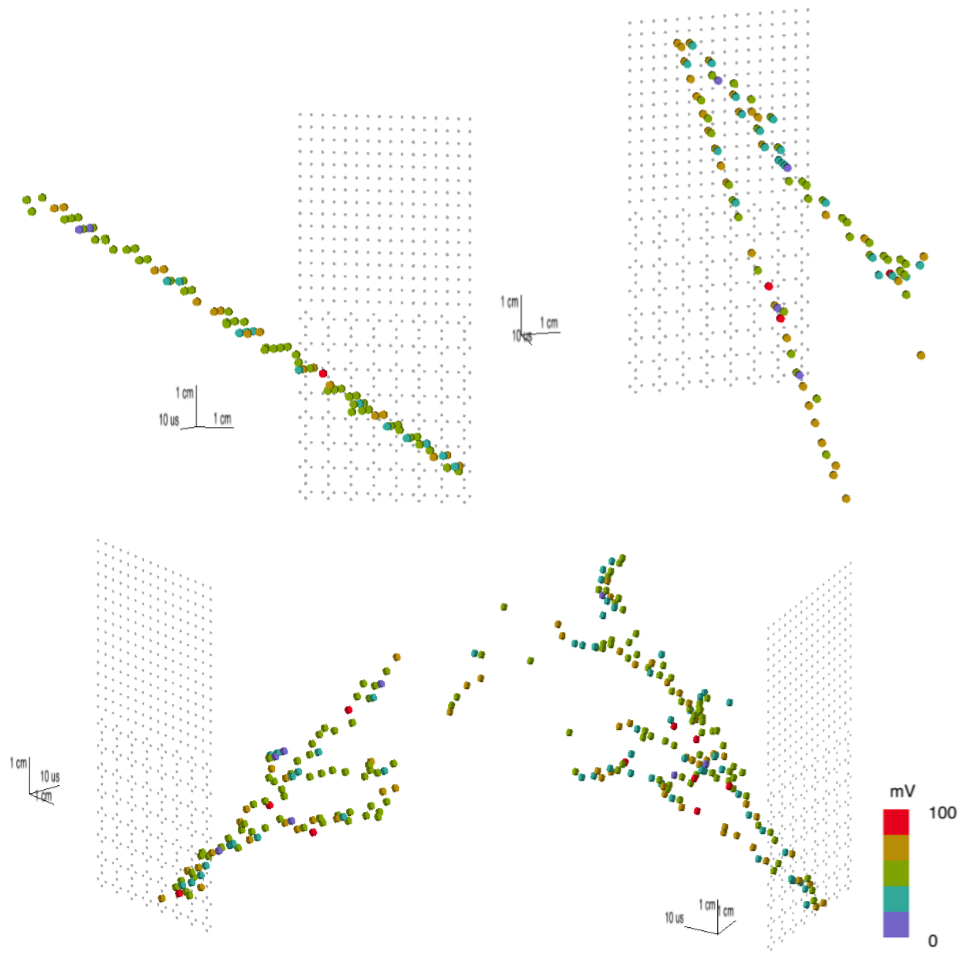}
  }
\caption{Four example 3D images of cosmic ray interactions detected
  using a 4.8~cm by 9.6~cm readout system in a 60-cm-drift LArTPC\@.
  The upper figures show relatively simple straight muon tracks, while
  the lower figures show small electromagnetic showers.  This is raw
  data collected from the system; no filtering or enhancement has been
  applied.  \hfill
\label{fig:tracks_60cm}}
\end{figure}

%
A 512-channel readout system was operated in a larger LArTPC system
located at the University~of~Bern\@.
The design of the LArTPC was similar, also 10~cm in diameter, but with
a drift length of 60~cm\@.
In addition to the larger active volume, this LArTPC had substantially
higher purity and therefore less electron loss.
Liquid argon purity was maintained by occasional recirculation of the
LAr through a purification system, enabling continuous operation over
multiple days.
The LArTPC was capable of operation at drift fields up to 1~kV/cm,
resulting in fewer electrons lost to prompt recombination.
A description of this LArTPC system is given in~\cite{Asaadi:2018oxk}.
The 512-channel readout system operated stably in this LArTPC for one
week, with roughly half of the data collected at a drift field of
500~V/cm and the other half at 1~kV/cm\@.
No degradation of readout performance was observed over this period.
Figure~\ref{fig:tracks_60cm} displays a few 3D images demonstrating the
typical signals of cosmic ray interactions observed during this test.
These images show raw data collected from the system; aside from
conversion of ADC value to voltage, no filtering, enhancement, or
other manipulation was applied.
The higher purity of the argon and increased drift field, relative to
the 10-cm-drift LArTPC, produced the more distinct 3D images shown in
figure~\ref{fig:tracks_60cm}\@.
See Supplemental Material at~\cite{SupplementalMaterial} for a 3D
animation of cosmic ray events.

During operation of these readout systems, a low rate ($\sim$1~Hz) of
high-amplitude pulses was observed on a fraction of channels using the
analog monitor of the ASICs\@.
This pulsing was found to originate from bubbling of the LAr, driven
by heat from the readout system located at the top face of the LArTPC.
At LBNL we subsequently rotated the entire LArTPC by 90 degrees, so
that the readout anode was now vertically-oriented and the electron
drift was horizontal.
The higher LAr head pressure and potentially improved convective
cooling for this orientation was sufficient to completely suppress
bubbling for both the 128-channel and 832-channel systems.
The 512-channel readout system was located at the top face of the
60-cm LArTPC at Bern, but it was possible to adjust the level of the
LAr in this cryostat.
Adjusting the LAr level so that the readout system was immersed at a
depth of at least $\sim$6~cm was sufficient to suppress bubbling in
this system.
During LArTPC testing we would occasionally collect data for the
specific purpose of monitoring the stability of the CSA output
pedestal voltage.
This was achieved by briefly enabling each channel with a
discriminator threshold set to the minimum value (0~V)\@.
The channel under test would self-trigger and digitize at the maximum
allowed rate, $\sim$500~kHz given the 5~MHz clock, and transmit ADC
values corresponding to zero input charge at the maximum 10~kHz rate
provided by the daisy chain communication.
Based on these tests, the CSA pedestal voltages were found to be
stable to within a few millivolts over the week of operation in the
Bern LArTPC system\@.
These data also served as a valuable test of the response to the
system to bursts of high rate operation and consequent higher heating,
far in excess of that expected from any reasonable LArTPC signals.
No bubbling was observed under these extreme conditions, except for
when the LAr level was very close to the 6~cm threshold.

\section{Discussion}\label{sec:discussion}

We have presented the design and results from operation of a low-power
readout system which provides true 3D imaging of particle interactions
in LArTPCs\@.
This readout system substantially bested the design targets for noise
(500~e$^-$~ENC) and power ($<$100~$\mu$W/channel), and also met the
requirements for cryogenic operation (87~K) and digital multiplexing
($>$1000 pads per data I/O channel).
The excellent performance of this first-generation system enabled
instrumentation and operation of LArTPCs with drift lengths of 10~cm
and 60~cm\@.
Signals from cosmic ray muons were imaged in 3D with no evidence of
spurious hits from noise, and required no data filtering or signal
enhancement.
The improved fidelity of true 3D readout overcomes issues of signal
ambiguities present for projective readout methods, which is one of
the factors currently hindering reconstruction of neutrino-nuclear
interactions at energies of a few GeV\@.
It should also enable LArTPC operation in a high-occupancy
environments, such as the DUNE Near Detector site.
Using pixelated 3D readout for LArTPCs has other potential advantages.
The rigid pixel PCBs are more mechanically robust and less sensitive
to microphonic noise.
Production of this readout relied on standard commercial electronics
fabrication techniques, and so production of large detector readout
can benefit from industrial economies of scale and QA/QC processes.
Assembly of large-scale readout from small ($<$0.25~m$^2$) modular
tiles should ease readout testing and installation relative to large
($>$10~m$^2$) wire-based readout planes.
Neither ASIC production costs ($\sim$\$2500/m$^{2}$) nor PCB
fabrication costs ($\sim$\$500/m$^{2}$) are prohibitive for
instrumentation of large detectors, although detailed methods for
low-cost assembly and testing of large-scale readout systems are still
in development.
This readout may also have potential for large directional dark matter
detectors that rely on negative ion drift for high-resolution imaging
of nuclear recoils~\cite{Battat:2016pap}.
The next steps for 3D readout development begin with a detailed
assessment of performance, including: gain stability, dynamic
response, cross-talk, and longevity, among other aspects.
It is also important to examine potential shortcomings of the current
design via a program of measurements in combination with simulation.
The use of a discriminator for signal timing adds slewing and other
signal shape dependence to the estimation of the signal time.
Simulation suggests this effect is generally small compared to the
current 200~ns timestamp precision, and much smaller than the
2.2~$\mu$s resample time, but a detailed assessment is still required.
The brief deadtime after each self-trigger will result in some charge
loss for signals of long duration, such as tracks with a steep angle
of incidence to the readout plane; a measurement of this effect using
cosmic ray data is in progress.
The impact of the loss of charge signals that are below the channel
threshold must be similarly examined.
The current daisy chain communication is not robust; the failure of
one ASIC renders the chain broken.
Revision of the current LArPix design would address some of the
shortcomings described above, allow for enhanced features, and help
address concerns about long-term cryogenic operation.
Design and testing of a larger modular readout panel for tiling large
surfaces would provide a natural step towards instrumenting large
LArTPCs\@.
%


\acknowledgments

We would like to thank our collaborators from the Laboratory for High
Energy Physics at the Univ.~of~Bern (Antonio Ereditato, James
Sinclair, Igor Kreslo, Damian Goeldi, among others) for their
pioneering work in pixel readout of LArTPCs, for extensive discussions
on readout design, for providing the 10-cm-drift LArTPC used for
testing at LBNL, and for operation of the 60-cm-drift LArTPC at
Bern\@.
Dan~McKinsey very kindly provided the high-purity liquid argon system
for hosting this LArTPC at LBNL, and Lucie Tvrznikova went out of her
way to help us adapt and operate this system.
The consistently useful suggestions from Yuan Mei at LBNL helped us
avoid a variety of pitfalls during both design and testing of these
readout systems.
This material is based upon work supported by the U.S. Department of
Energy, Office of Science, Office of High Energy Physics, under
contract number DE-AC02-05CH11231\@.


\bibliographystyle{unsrt}
\bibliography{DUNE}

\begin{thebibliography}{10}

\bibitem{Nygren:1978rx}
J.~Marx and D.~Nygren,
\newblock {\em The Time Projection Chamber},
\newblock \href{http://dx.doi.org/10.1063/1.2994775}{{\em Phys. Today} {\bf 31} (1978) 46}.

\bibitem{Olive:2016xmw}
C.~Patrignani et~al,
\newblock {\em Review of Particle Physics},
\newblock \href{http://dx.doi.org/}{{\em Chin. Phys.} {\bf C40} (2016) 100001}.

\bibitem{Rubbia:1977zz}
C.~Rubbia,
\newblock {\em The Liquid Argon Time Projection Chamber: A New Concept for Neutrino Detectors},
\newblock {\em CERN-EP-INT-77-08} (1977).

\bibitem{Chen:1976pp}
H.~H. Chen, P.~E. Condon, B.~C. Barish, and F.~J. Sciulli,
\newblock {\em A Neutrino detector sensitive to rare processes. I. A Study of
  neutrino electron reactions},
\newblock {\em FERMILAB-PROPOSAL-0496} (1976).

\bibitem{Willis:1974gi}
W.~J. Willis and V.~Radeka,
\newblock {\em Liquid Argon Ionization Chambers as Total Absorption Detectors},
\newblock \href{https://doi.org/10.1016/0029-554X(74)90039-1}{{\em Nucl. Instrum. Meth.} {\bf 120} (1974) 221}.

\bibitem{Amerio:2004ze}
S.~Amerio et~al,
\newblock {\em Design, construction and tests of the ICARUS T600 detector},
\newblock \href{http://dx.doi.org/10.1016/j.nima.2004.02.044}{{\em Nucl. Instrum. Meth.} {\bf A527} (2004) 329}.

\bibitem{Anderson:2012vc}
C.~Anderson et~al,
\newblock {\em The ArgoNeuT Detector in the NuMI Low-Energy beam line at Fermilab},
\newblock \href{http://dx.doi.org/10.1088/1748-0221/7/10/P10019}{{\em JINST} {\bf 7} (2012) 10019}.

\bibitem{Acciarri:2016smi}
R.~Acciarri et~al,
\newblock {\em Design and Construction of the MicroBooNE Detector},
\newblock \href{http://dx.doi.org/10.1088/1748-0221/12/02/P02017}{{\em JINST} {\bf 12} (2017) 02017}.

\bibitem{Antonello:2015lea}
M.~Antonello et~al,
\newblock {\em A Proposal for a Three Detector Short-Baseline Neutrino Oscillation
  Program in the Fermilab Booster Neutrino Beam},
\newblock \href{http://arxiv.org/abs/1503.01520}{arXiv:1503.01520}.

\bibitem{Acciarri:2016crz}
R.~Acciarri et~al,
\newblock {\em Long-Baseline Neutrino Facility (LBNF) and Deep Underground Neutrino
  Experiment (DUNE)},
\newblock \href{http://arxiv.org/abs/1601.05471}{arXiv:1601.05471}.

\bibitem{Acciarri:2017sde}
R.~Acciarri et~al,
\newblock {\em Noise Characterization and Filtering in the MicroBooNE Liquid Argon
  TPC},
\newblock \href{http://dx.doi.org/10.1088/1748-0221/12/08/P08003}{{\em JINST} {\bf 12} (2017) 08003}.

\bibitem{Adams:2018dra}
C.~Adams et~al,
\newblock {\em Ionization Electron Signal Processing in Single Phase LArTPCs I.
  Algorithm Description and Quantitative Evaluation with MicroBooNE
  Simulation},
\newblock \href{http://dx.doi.org/10.1088/1748-0221/13/07/P07006}{{\em JINST} {\bf 13} (2018) 07006}.

\bibitem{Adams:2018gbi}
C.~Adams et~al,
\newblock {\em Ionization Electron Signal Processing in Single Phase LArTPCs II.
  Data/Simulation Comparison and Performance in MicroBooNE},
\newblock \href{http://dx.doi.org/10.1088/1748-0221/13/07/P07007}{{\em JINST} {\bf 13} (2018) 07007}.

\bibitem{Qian:2018qbv}
X.~Qian, C.~Zhang, B.~Viren, and M.~Diwan,
\newblock {\em Three-dimensional Imaging for Large LArTPCs},
\newblock \href{https://doi.org/10.1088/1748-0221/13/05/P05032}{{\em JINST} {\bf 13} (2018) 05032}.

\bibitem{Alme:2010ke}
J.~Alme et~al,
\newblock {\em The ALICE TPC, a large 3-dimensional tracking device with fast
  readout for ultra-high multiplicity events},
\newblock \href{http://dx.doi.org/10.1016/j.nima.2010.04.042}{{\em Nucl. Instrum. Meth.} {\bf A622} (2010) 316}.

\bibitem{Anderson:2003ur}
M.~Anderson et~al,
\newblock {\em The Star time projection chamber: A Unique tool for studying high
  multiplicity events at RHIC},
\newblock \href{http://dx.doi.org/10.1016/S0168-9002(02)01964-2}{{\em Nucl. Instrum. Meth.} {\bf A499} (2003) 659}.

\bibitem{Asaadi:2018oxk}
J.~Asaadi et~al,
\newblock {\em First Demonstration of a Pixelated Charge Readout for Single-Phase
  Liquid Argon Time Projection Chambers},
\newblock \href{http://arxiv.org/abs/1801.08884}{arXiv:1801.08884}.

\bibitem{Miyajima:1974zz}
M.~Miyajima, T.~Takahashi, S.~Konno, T.~Hamada, S.~Kubota, H.~Shibamura, and
  T.~Doke,
\newblock {\em Average energy expended per ion pair in liquid argon},
\newblock \href{http://dx.doi.org/10.1103/PhysRevA.9.1438}{{\em Phys. Rev.} {\bf A9} (1974) 1438}.

\bibitem{PhysRevA.10.1452}
M.~Miyajima, T.~Takahashi, S.~Konno, T.~Hamada, S.~Kubota, H.~Shibamura, and
  T.~Doke,
\newblock {\em Erratum: Average energy expended per ion pair in liquid argon},
\newblock \href{http://dx.doi.org/10.1103/PhysRevA.10.1452}{{\em Phys. Rev.} {\bf A10} (1974) 1452}.

\bibitem{Amoruso:2004dy}
S.~Amoruso et~al,
\newblock {\em Study of electron recombination in liquid argon with the ICARUS
  TPC},
\newblock \href{http://dx.doi.org/10.1016/j.nima.2003.11.423}{{\em Nucl. Instrum. Meth.} {\bf A523} (2004) 275}.

\bibitem{Acciarri:2013met}
R.~Acciarri et~al,
\newblock {\em A study of electron recombination using highly ionizing particles in
  the ArgoNeuT Liquid Argon TPC},
\newblock \href{http://dx.doi.org/10.1088/1748-0221/8/08/P08005}{{\em JINST} {\bf 8} (2013) 08005}.

\bibitem{Antonello:2014eha}
M.~Antonello et~al,
\newblock {\em Experimental observation of an extremely high electron lifetime with
  the ICARUS-T600 LAr-TPC},
\newblock \href{http://dx.doi.org/10.1088/1748-0221/9/12/P12006}{{\em JINST} {\bf 9} (2014) 12006}.

\bibitem{Acciarri:2016sli}
R.~Acciarri et~al,
\newblock {\em First Observation of Low Energy Electron Neutrinos in a Liquid Argon
  Time Projection Chamber},
\newblock \href{http://dx.doi.org/10.1103/PhysRevD.95.072005}{{\em Phys. Rev.} {\bf D95} (2017) 072005}.

\bibitem{Walkowiak:2000wf}
W.~Walkowiak,
\newblock {\em Drift velocity of free electrons in liquid argon},
\newblock \href{http://dx.doi.org/10.1016/S0168-9002(99)01301-7}{{\em Nucl. Instrum. Meth.} {\bf A449} (2000) 288}.

\bibitem{LuciesThesis}
L.~Tvrznikova,
\newblock Ph.D Thesis (in preparation), Yale University 2018.

\bibitem{SupplementalMaterial}
\newblock \url{http://arxiv.org/src/1808.02969/anc/larpix_cosmics_60cmTPC.mp4}.

\bibitem{Battat:2016pap}
J.~B.~R. Battat et~al,
\newblock {\em Readout technologies for directional WIMP Dark Matter detection},
\newblock \href{http://dx.doi.org/10.1016/j.physrep.2016.10.001}{{\em Phys. Rept.} {\bf 662} (2016) 1}.


\end{thebibliography}







\end{document}